\documentclass[reprint,superscriptaddress,preprintnumbers,amsmath,amssymb,aps,prd,tightenlines,longbibliography,nofootinbib]{revtex4-2}

\usepackage{graphicx}
\usepackage[dvipsnames]{xcolor}
\usepackage{amsmath,amssymb,bm,subdepth}
\usepackage[colorlinks=true
,urlcolor=blue
,anchorcolor=blue
,citecolor=blue
,filecolor=blue
,linkcolor=red
,menucolor=blue
,linktocpage=true
,pdfproducer=medialab
,pdfa=true
]{hyperref}
\usepackage{mathtools}

\newcommand{\be}{\begin{equation}}
\newcommand{\ee}{\end{equation}}
\newcommand{\bea}{\begin{equation}\begin{aligned}}
\newcommand{\eea}{\end{aligned}\end{equation}}

\newcommand{\bx}{\mathbf{x}}
\newcommand{\bp}{\mathbf{p}}
\newcommand{\bq}{\mathbf{q}}
\newcommand{\ga}{g_{a\gamma\gamma}}

\newcommand{\la}{\langle}
\newcommand{\ra}{\rangle}
\DeclarePairedDelimiter\ket{|}{\ra} 
\DeclarePairedDelimiter\bra{\la}{|}
\newcommand{\normord}[1]{\,:\mathrel{#1}:\,}

\newcommand{\DM}{{\scriptscriptstyle \textrm{DM}}}

\makeatletter
\newcounter{savesection}
\newcounter{apdxsection}
\renewcommand\appendix{\par
  \setcounter{savesection}{\value{section}}%
  \setcounter{section}{\value{apdxsection}}%
  \setcounter{subsection}{0}%
  \gdef\thesection{\@Alph\c@section}}
\newcommand\unappendix{\par
  \setcounter{apdxsection}{\value{section}}%
  \setcounter{section}{\value{savesection}}%
  \setcounter{subsection}{0}%
  \gdef\thesection{\@arabic\c@section}}
\makeatother

\begin{document}

\title{Intrinsically Quantum Effects of Axion Dark Matter are Undetectable}
 

\author{Yunjia Bao}
\affiliation{Enrico Fermi Institute, The University of Chicago, Chicago, IL 60637, USA}
\affiliation{Department of Physics, The University of Chicago, Chicago, IL 60637, USA}
\affiliation{Leinweber Institute for Theoretical Physics, The University of Chicago, Chicago,
IL 60637, USA}
\affiliation{Kavli Institute for Cosmological Physics, The University of Chicago, Chicago,
IL 60637, USA}

\author{Dhong Yeon Cheong}
\affiliation{Enrico Fermi Institute, The University of Chicago, Chicago, IL 60637, USA}
\affiliation{Department of Physics, The University of Chicago, Chicago, IL 60637, USA}
\affiliation{Leinweber Institute for Theoretical Physics, The University of Chicago, Chicago,
IL 60637, USA}
\affiliation{Kavli Institute for Cosmological Physics, The University of Chicago, Chicago,
IL 60637, USA}
\affiliation{Department of Physics and IPAP, Yonsei University, Seoul 03722, Republic of Korea}

\author{Nicholas L.~Rodd}
\affiliation{Theory Group, Lawrence Berkeley National Laboratory, Berkeley, CA 94720, USA}
\affiliation{Leinweber Institute for Theoretical Physics, University of California, Berkeley, CA 94720, USA}

\author{Joey Takach}
\affiliation{Theory Group, Lawrence Berkeley National Laboratory, Berkeley, CA 94720, USA}
\affiliation{Leinweber Institute for Theoretical Physics, University of California, Berkeley, CA 94720, USA}

\author{Lian-Tao Wang}
\affiliation{Enrico Fermi Institute, The University of Chicago, Chicago, IL 60637, USA}
\affiliation{Department of Physics, The University of Chicago, Chicago, IL 60637, USA}
\affiliation{Leinweber Institute for Theoretical Physics, The University of Chicago, Chicago,
IL 60637, USA}
\affiliation{Kavli Institute for Cosmological Physics, The University of Chicago, Chicago,
IL 60637, USA}

\author{Kevin Zhou}
\affiliation{Theory Group, Lawrence Berkeley National Laboratory, Berkeley, CA 94720, USA}
\affiliation{Leinweber Institute for Theoretical Physics, University of California, Berkeley, CA 94720, USA}

\begin{abstract}
Is the usual treatment of axion dark matter as a classical field reliable?
We show that the answer is subtle: the axion field could well be in a quantum state that has no complete classical description, but realistic detectors cannot tell the difference.
To see this, we solve a fully quantum model of axion detection using quantum optics techniques.
We show that intrinsically quantum effects are washed out by mode averaging or small amounts of noise, and significantly suppressed by the weakness of the axion coupling.
Our work exemplifies that there should always be a classical analog for axion dark matter effects, extends to other wave (ultralight) dark-matter candidates, and gives a general method to compute the effects of exotic dark-matter states.
\end{abstract}

\maketitle


Quantum measurement techniques are poised to become an indispensable tool in the search for wave dark matter (DM)~\cite{Adams:2022pbo,Antypas:2022asj,Berlin:2024pzi,Fang:2024ple}.
There are multiple paths beyond the standard quantum limit.
It can be evaded with squeezed states, as pioneered by LIGO for gravitational waves~\cite{LIGOScientific:2011imx,LIGOScientific:2013pcc} and HAYSTAC for DM~\cite{Malnou:2018dxn,HAYSTAC:2020kwv,HAYSTAC:2023cam,HAYSTAC:2024jch}, or circumvented entirely with photon number measurements~\cite{Lamoreaux:2013koa}.
In the drive for DM, groups are developing microwave photon counters sensitive to single photons exiting a detector cavity~\cite{Lescanne:2020awk,Balembois:2023bvs,Braggio:2024xed,pankratov2022towards,Pankratov:2024kdv,Pankratov:2025cby,Kuo:2024duq}, and techniques to measure the photons inside a cavity with transmon qubits~\cite{Chakram:2021bxb,Dixit:2020ymh,Gu:2025pms}, which can also be used to prepare the cavity in nonclassical Fock~\cite{Agrawal:2023umy} or cat~\cite{Zheng:2025qgv} states.
Even more ambitious proposals involve two-mode squeezing~\cite{Wurtz:2021cnm,Jiang:2022vpm,Shi:2022wpf,Shi:2024qhd} and entanglement of multiple cavities or qubits~\cite{Brady:2022bus,Chen:2022quj,Chen:2023swh,Ito:2023zhp,Chen:2024aya,Shu:2024nmc,Fukuda:2025zcf}.

Despite operating in a highly quantum regime, these efforts all treat wave DM as a classical field.
This is because it is a bosonic quantum field with ultralight mass, $m_\DM \ll 10 \, \mathrm{eV}$, which implies an enormous number of quanta per field mode; assuming standard virialization, it is ${\sim}\rho_\DM / (m_\DM \Delta p_\DM^3) \sim 10^{28} (\mu{\rm eV}/m_\DM)^4$.
But high occupation does not automatically imply DM behaves classically~\cite{Cheong:2024ose}: cat or squeezed states are inherently nonclassical but can have arbitrarily high occupancy.
In fact, wave DM could be produced in a nonclassical state or evolve into one in the galaxy, while avoiding decoherence due to its weak coupling.
This raises the question of whether there are fundamentally new signatures of DM that are missed by the classical field approximation.

In this Letter, we construct a fully quantum description of the interaction of wave DM with a detector, and find two serious obstacles to detecting intrinsically quantum effects.
First, realistic detectors couple to \textit{effective} DM modes, which aggregate many DM modes.
Through a quantum analog of the central limit theorem, this tends to wash out quantum effects, even if the individual modes are in nonclassical states.
Second, intrinsically quantum signatures are suppressed by the low efficiency $\eta$ of DM-photon conversion.
Even under optimistic assumptions, they can easily be overwhelmed by detector noise, and even if not would require an extraordinarily long time to resolve.
We focus on axion DM detection in a cavity haloscope~\cite{Sikivie:1983ip}, though the argument extends broadly to other weakly coupled signals.
Further details are provided in a pair of accompanying works~\cite{Bao:2026prd,Bao:2026kbx}.

To summarize, though we may be poised to detect axion DM, which could be in a highly quantum state, our instruments are not capable of resolving nonclassical effects in the foreseeable future.
Though some works have claimed otherwise, we show axion DM can always be treated as a classical field, possibly with an exotic energy or fluctuation distribution.
More generally, experiments can safely proceed by treating wave DM as classical.

\vspace{0.2cm}
\noindent {\bf The Quantum State of DM.}
%
Our starting point is the seminal insight of Glauber~\cite{Glauber:1963tx} and Sudarshan~\cite{Sudarshan:1963ts} that a quantum field mode in a coherent state $\ket{\alpha}$ acts on matter in the same way as a classical field mode with complex amplitude $\alpha$. 
Thus, any mixture of coherent states, described by the density matrix
\be
\hat{\rho} = \int d\alpha \, P(\alpha) \, \ket{\alpha} \bra{\alpha}
\label{eq:GSP}
\ee
for $P(\alpha) \geq 0$, acts like a probabilistic mixture of classical field values. 
(Here $d \alpha = d \textrm{Re}\alpha\, d \textrm{Im}\alpha$.)
Nonclassical states can also be described by Eq.~\eqref{eq:GSP}, but involve $P$-functions which attain negative values. 
Examples include squeezed states, Fock states, and cat states that are~superpositions of coherent states $\ket{\alpha}$ and $\ket{\beta}$ with $|\alpha - \beta| \gg 1$.

Turning to the axion, misalignment production begins with a relatively well-defined field value, so it is often said to be in a coherent state.
We are not aware of a proper justification of this claim; regardless, the pre-inflationary axion has its state subsequently squeezed during inflation, e.g.~Refs.~\cite{Kuss:2021gig,Kalia:2025kmr}, whereas for post-inflationary production the contribution from the axion string network, e.g.~Refs.~\cite{Gorghetto:2024vnp,Saikawa:2024bta,Kim:2024wku,Benabou:2024msj}, is not a pure coherent state.
Furthermore, during the collapse and subsequent evolution of our galaxy, the axion state should change substantially due to gravitational interactions.
In the end, the state of DM in the laboratory is described by a joint $P$-function over the modes of the field,
\be
\hat{\rho}_\DM = \int d\bm{\alpha} \, P_\DM(\bm{\alpha}) \, \ket{\bm{\alpha}}\bra{\bm{\alpha}}
\ee
where $\bm{\alpha} = (\alpha_1, \alpha_2, \ldots)$ denotes the DM modes, and again the system can be described classically if $P_\DM \geq 0$.

Simulations indicate that quantum corrections generically grow exponentially during halo collapse, but are simultaneously suppressed by decoherence~\cite{Eberhardt:2022rcp,Eberhardt:2023axk}.
However, cat states with $|\alpha| = |\beta|$ do not decohere rapidly because both branches of the wavefunction have the same average gravitational potential~\cite{Allali:2020ttz,Allali:2020shm,Allali:2021puy}.
Further, there is an extensive debate over whether the axion evolves to a state where only one mode is occupied, which would drastically increase its coherence time and could be a description of a Bose--Einstein condensate; see e.g. Refs.~\cite{Semikoz:1995rd,Sikivie:2009qn,Guth:2014hsa,Marsh:2015xka,Budker:2023sex}.

Identifying the quantum state of local axion DM is evidently a highly nontrivial question, which we do not aim to resolve here.
Instead, we remain agnostic as to the form of $\hat{\rho}_\DM$ and determine the requirements for a detector to observe a genuinely quantum effect.

\vspace{0.2cm}
\noindent {\bf A Model of DM Detection.}
%
Cavity haloscopes search for axion DM, $\phi$, coupled to photons through $\mathcal{L}_{\mathrm{int}} = \ga \phi\,\mathbf{E} \cdot \mathbf{B}$.
The DM resonantly converts to photons within the cavity volume $V_c$ and background magnetic field $B_0 \hat{\mathbf{z}}$.
The Hamiltonian describing the interaction is
\be
H_{\mathrm{int}} = -\ga \, B_0 \int_{V_c} \hspace{-0.15cm} d^3 \bx \, \phi(\bx) E_z(\bx).
\ee
The fields have the following mode expansion,
\bea
E_z(\bx) &= \sum_\ell \sqrt{\frac{\omega_\ell}{2}} \, \Big(i c_\ell \tilde{E}^*_{\ell,z}(\bx) + \mathrm{h.c.} \Big) \\
\phi(\bx) &= \sum_{\bp} \frac{1}{\sqrt{2 \omega_p \mathcal{V}}} \, \Big(a_{\bp} e^{i \bp \cdot \bx} + \mathrm{h.c.} \Big)
\eea
where $\omega_\ell$ is the cavity mode angular frequency, $\tilde{E}_{\ell, z}$ is the $z$-component of the mode profile normalized to $\int_{V_c} \!d^3\bx \, |\tilde{\mathbf{E}}_\ell|^2 = 1$, $\omega_p = m_\DM + K_p$ is the axion energy (with $K_p$ the kinetic energy), and $\mathcal{V}$ is a fiducial axion quantization volume.

We suppose a cavity mode is on resonance with the axion field, $\omega_\ell \simeq m_\DM$, and neglect the modes off resonance.
Working in the interaction picture, applying the rotating wave approximation, and suppressing the resonant mode subscript, the Hamiltonian becomes
\be
H_{\mathrm{int}}(t) \simeq \frac{1}{2}\ga B_0 \sqrt{\frac{V_c}{\mathcal{V}}} \, \Big(i c^\dagger \sum_{\bp} C_{\bp} e^{- i K_p t} a_{\bp} + \mathrm{h.c.} \Big)
\label{eq:Hint-full}
\ee
when written in terms of a dimensionless overlap factor
\be
C_{\bp} = \sqrt{\frac{m_\DM}{\omega_p \, V_c}} \int_{V_c} \hspace{-0.15cm}d^3 \bx \, e^{i \bp \cdot \bx} \tilde{E}_z(\bx)
\ee
that generalizes the cavity form factor to arbitrary DM momenta; cf. Ref.~\cite{Dror:2021nyr}.
The conventional form factor can be identified as $|C_{\bm{0}}|^2$.

Equation~\eqref{eq:Hint-full} reveals that the cavity couples to an \textit{effective} axion mode, defined as
\be
a_{\mathrm{eff}}(t) \equiv \frac{1}{2} \sqrt{\frac{V_c}{\Omega {\cal V}}} \sum_{\bp} C_{\bp} e^{- i K_p t} a_{\bp}.
\label{eq:aeff_def}
\ee
An analogous reorganization of modes is used for interferometers~\cite{Ley01041986,Hariharan01011993} and the double slit experiment~\cite{10.1119/1.10857,PhysRevLett.134.133603}.
Using this, the Hamiltonian simplifies considerably to
\be
H_{\mathrm{int}}(t) \simeq i g \, \big(c^\dagger a_{\mathrm{eff}}(t) - c \, a_{\mathrm{eff}}^\dagger(t) \big)
\label{eq:Hint_def}
\ee
where $g = \ga B_0 \sqrt{\Omega}$ is a coupling with units of frequency.
We normalize $a_{\mathrm{eff}}$ with $[a_{\mathrm{eff}}, a_{\mathrm{eff}}^\dagger] = 1$, which then determines $\Omega \sim |C_{\bm{0}}|^2$.
The effective mode's mean occupancy is equal to the expected number of axions in the cavity volume, $N_{\mathrm{eff}} = \la a_{\mathrm{eff}}^\dagger a_{\mathrm{eff}} \ra \simeq \rho_\DM V_c / m_\DM$.
These results are justified in the End Matter, and further discussion of effective modes is provided in Ref.~\cite{Bao:2026prd}.

Many haloscopes continuously monitor the cavity through a port with coupling rate $1/t_m$, where $t_m$ is the effective measurement interval.
This can be treated with input-output theory, as reviewed in Refs.~\cite{Clerk:2008tlb,Malnou:2018dxn,Beckey:2023shi,Bernal:2024hcc}.
We discuss that case in Ref.~\cite{Bao:2026kbx}.
Here we take the simplifying case where information is read out by repeatedly preparing and projectively measuring the cavity after each time $t_m$.
Critical coupling roughly corresponds to setting $t_m$ to the cavity decay time $Q_c / m_\DM$, where $Q_c$ is the cavity's intrinsic quality factor, and we optimistically neglect cavity damping on this timescale.
We adopt this model below.

Within an axion coherence time $\tau_\DM \sim 1/\Delta \omega$, with $\Delta \omega$ the energy width of the occupied DM modes, the phase factors $e^{- i K_p t}$ in Eq.~\eqref{eq:aeff_def} do not substantially change relative to each other.
(This is essentially by definition: the coherence time specifies how long it takes the different modes to develop an $\sim$$2\pi$ phase relative to one another; see Ref.~\cite{Cheong:2024ose} for further discussion.)
Then, the effective mode can be regarded as having a fixed state, with the following density matrix weighting the different values,
\be
\hat{\rho}^{\textrm{eff}}_\DM = \int d\bm{\alpha}\,P_\DM(\bm{\alpha})\, \ket{\alpha_{\mathrm{eff}}(t_0)} \bra{\alpha_{\mathrm{eff}}(t_0)}
\ee
where $a_{\mathrm{eff}}(t_0) \ket{\bm{\alpha}} = \alpha_{\mathrm{eff}}(t_0) \ket{\bm{\alpha}}$, and $t_0$ is the initial time.
From this density matrix we can infer that the effective mode is controlled by the following $P$-function,
\be
P^{\mathrm{eff}}_\DM(\beta) = \int d\bm{\alpha} \, P_\DM(\bm{\alpha})\, \delta \big(\beta - \alpha_{\mathrm{eff}}(t_0) \big).
\label{eq:Peff_def}
\ee
Measurements spaced out over a time $t_m \gtrsim \tau_\DM$ would involve a time-varying effective mode, or equivalently would require multiple effective modes to describe.
This variation tends to suppress the visibility of nonclassical effects, so we optimistically assume $t_m \lesssim \tau_\DM$, during which the effective mode's state is fixed.

So far, no assumption regarding the axion state has been made.
Assuming standard virialization, one could argue that the DM modes should become independent, so that the state factorizes as $P_\DM(\bm{\alpha}) = \prod_j P^j_\DM(\alpha_j)$ (cf. Refs.~\cite{Foster:2017hbq,Cheong:2024ose}).
In that case, Eq.~\eqref{eq:Peff_def} reduces to a convolution over all $P^j_\DM$, weighted by the coupling between the mode and the detector.
Then by the (quantum) central limit theorem~\cite{Glauber:1963tx} (reviewed in detail in Ref.~\cite{Bao:2026prd}), $P^{\mathrm{eff}}_\DM(\alpha)$ becomes a Gaussian, and therefore effectively classical.
This remains true even if one can resolve the axion linewidth, as one can define an effective mode for each frequency bin.
The argument holds until the fundamental modes of the field are resolved.

This justifies the common assumption that the axion as seen by a detector is described by a classical Gaussian random field~\cite{Foster:2017hbq,Cheong:2024ose}.
To see nonclassical effects, it is insufficient for individual modes to be nonclassical.
Instead, correlations between the modes must be strong enough to cause the central limit theorem to fail.
This could occur, e.g.~if the axion condenses into a single mode, so we let $P^{\mathrm{eff}}_\DM(\alpha)$ be arbitrarily nonclassical below.

\vspace{0.2cm}
\noindent {\bf Washing Out Quantum Effects.}
%
The model in Eq.~\eqref{eq:Hint_def} can be solved exactly.
To see this, note $H_{\mathrm{int}}$ rotates the modes into one another by an angle $gt$, e.g.
\be
e^{i H_{\mathrm{int}} t} \, c \, e^{- i H_{\mathrm{int}} t} = c \cos gt + a_{\mathrm{eff}} \sin gt.
\label{eq:conj_result}
\ee
Accordingly, an initial interaction picture coherent state $\ket{\alpha}_\DM \ket{\beta}_{\mathrm{cav}}$ evolves to $\ket{\alpha \cos gt - \beta \sin gt}_\DM \ket{\beta \cos gt + \alpha \sin gt}_{\mathrm{cav}}$.
This implies (see the End Matter) that if the cavity state $P$-function is initially $P_{\mathrm{cav}}^i(\alpha)$, after a time $t_m$ it evolves to
\be 
\hspace{-0.1cm}P_{\mathrm{cav}}^f(\alpha) =\!\int\!d \beta \, \frac{P^{\mathrm{eff}}_\DM(\beta / \sqrt{\eta})}{\eta} \frac{P_{\mathrm{cav}}^i\big((\alpha - \beta)/\sqrt{1-\eta}\big)}{1-\eta}
\label{eq:final_Pcav}
\ee
which is simply a scaled convolution of the $P$-functions for the effective axion mode and the initial cavity mode.
Above, the very small conversion efficiency is
\bea
\eta &= \sin^2(g t_m) \\
&\sim 10^{-22} \left( \frac{\ga}{10^{-15} \, \mathrm{GeV}^{-1}} \frac{B_0}{10 \, \mathrm{T}^{\vrule height 4.5pt width 0pt}} \frac{Q_c}{10^5} \frac{10^{-5} \, \mathrm{eV}}{m_\DM} \right)^2 
\label{eq:eta_value}
\eea
where we took $C_{\bm{0}} \sim 1$ and normalized to typical values for cavity haloscopes.
Equation~\eqref{eq:final_Pcav} is our main result.
It provides a direct path to testing whether DM can produce observable nonclassical signatures in a haloscope.

At first glance, it may seem that nonclassical effects are easy to observe.
If the cavity is perfectly initialized in the vacuum state, $P_{\mathrm{cav}}^i(\alpha) = \delta(\alpha)$, then the final cavity state is related to the DM state by scaling,
\be
P_{\mathrm{cav}}^f(\alpha) = \frac{P^{\mathrm{eff}}_\DM(\alpha / \sqrt{\eta})}{\eta}. 
\label{eq:Pcav_ideal}
\ee
Thus, any negativity in the DM $P$-function is imprinted on the cavity mode, driving it into a nonclassical state.
Yet as we show below, the minute value of $\eta$ proves a key obstruction to observing the nonclassicality.

\begin{figure}
\vspace{-0.2cm}
\centering
\includegraphics[width=\columnwidth]{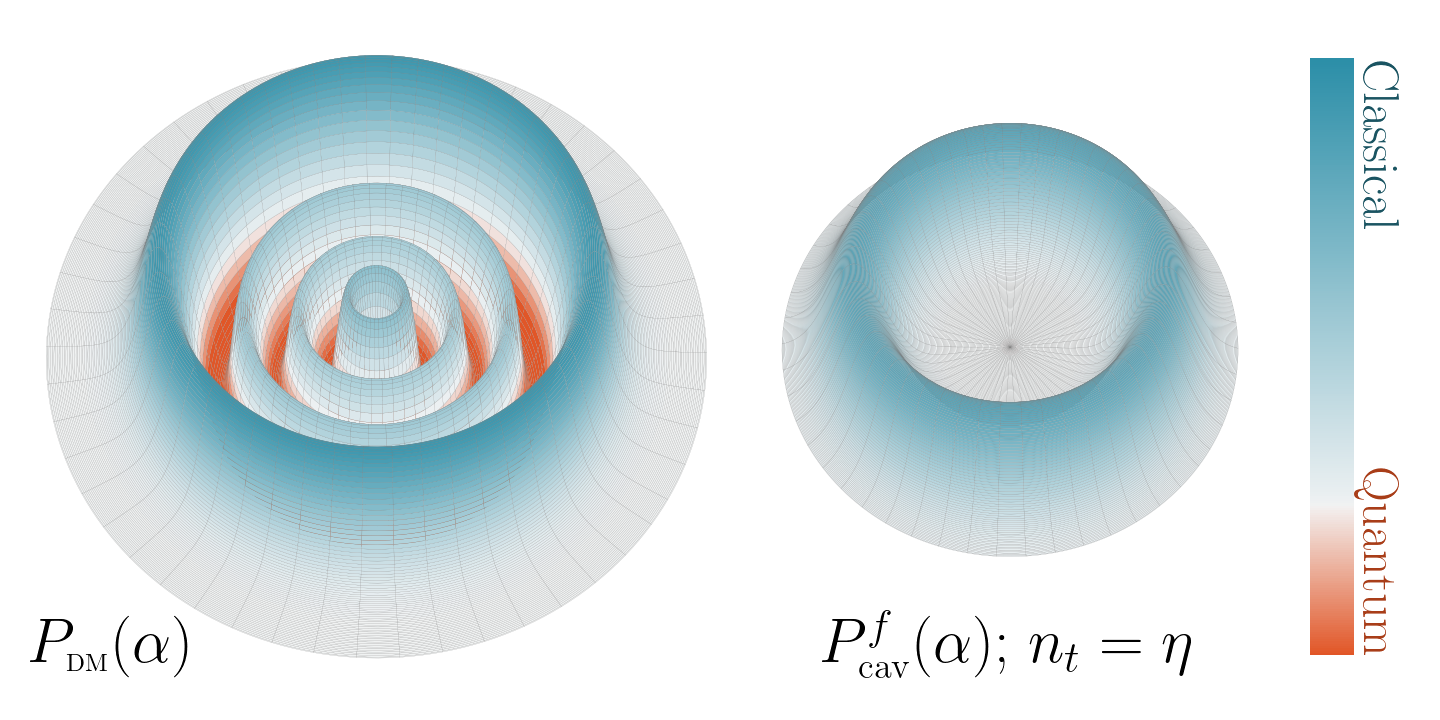}
\vspace{-0.8cm}
\caption{If the DM $P$-function (left) is imprinted on a noise-free cavity, its negativity can be preserved.
However, it can be washed out by an extremely small amount of thermal noise (right), yielding measurement statistics equivalent to a classical ensemble.
The DM $P$-function shown is a Fock state with Gaussian noise; see the End Matter for details.}
\vspace{-0.5cm}
\label{fig:3d_plot}
\end{figure}

Moreover, in practice, the cavity cannot be perfectly prepared in the vacuum state.
Instead, it always carries a Gaussian spread due to thermal excitation~\cite{carmichael2013statistical},
\be
P_{\mathrm{cav}}^i(\alpha) = \frac{e^{-|\alpha|^2 / n_t}}{\pi \, n_t}
\label{eq:P_initial_thermal}
\ee
with $n_t$ the mean thermal occupancy.
If $n_t = 1$, then convolving this Gaussian with any other $P$-function would yield the Husimi function, which is nonnegative~\cite{mandel1995optical}.
In our case, the DM $P$-function begins scaled down by a factor of $\sqrt{\eta}$, so its negativity survives only under the more stringent condition $n_t \lesssim \eta$, which follows directly from Eqs.~\eqref{eq:final_Pcav} and \eqref{eq:P_initial_thermal}.
This corresponds to requiring a temperature 
\be
T \lesssim \frac{m_\DM}{\log(1/\eta)} \simeq 2 \, \mathrm{mK} \left( \frac{m_\DM}{10^{-5} \, \mathrm{eV}} \right)
\label{eq:T_requirement}
\ee
where we used $n_t \simeq \exp(-m_\DM/T)$, valid for $T \lesssim m_\DM$, and the efficiency in Eq.~\eqref{eq:eta_value}.

At microwave frequencies and below, this is a stringent requirement on the physical temperature, and it also applies to all other sources of noise in the readout chain.
Therefore, for most haloscopes, nonclassical effects are already washed out before measurement, as we illustrate in Fig.~\ref{fig:3d_plot}, though this might be avoided for haloscopes targeting $m_\DM \gtrsim \mathrm{meV}$.
In principle, for temperatures exceeding this bound, one could still infer negativity of $P_\DM^{\mathrm{eff}}$ if $n_t$ was known accurately.
However, as shown in the End Matter, this requires determining $n_t$ to ${\sim} \eta$ precision, which is highly unrealistic.
We further show there that for $m_\DM \ll \mu\mathrm{eV}$, detection of quantum effects remains very difficult, cf. Eq.~\eqref{eq:eta_value}.

To our knowledge, this argument has not appeared in the quantum optics literature, because there one rarely considers $\eta \ll 1$.
The closest analogue we have found is an argument that light amplifiers can remove nonclassical effects by injecting Gaussian noise~\cite{Hong:85,KLYSHKO1989334,PhysRevA.47.3160}.

\vspace{0.2cm}
\noindent {\bf Suppression of Nonclassicality Measures.}
%
Let us optimistically assume that the cavity is prepared in the vacuum state and the DM has an infinite coherence time.
Even then, nonclassical effects remain exceptionally hard to observe.
As we quantify in examples below, doing so would require an excessively long experimental integration time, $t_{\mathrm{int}}$.
The challenge originates from Eq.~\eqref{eq:Pcav_ideal}.
Because $P^{\mathrm{eff}}_\DM$ can be negative in regions of at most ${\sim} 1$ in size (as smearing any $P$-function by a unit Gaussian gives the nonnegative Husimi function), $P_{\mathrm{cav}}^f$ inherits negativity over regions of size ${\sim} \sqrt{\eta}$, and thus rapidly oscillates in sign.
Since all observables can be expressed as integrals over $P_{\mathrm{cav}}^f$, nonclassical effects are generically suppressed.

The probability distribution for measurements of the quadrature operator $X = (c+c^\dagger)/\sqrt{2}$ is
\be
p(x) = \int d \alpha \, P^{\mathrm{eff}}_\DM(\alpha) \, |\la x | \sqrt{\eta} \, \alpha \ra|^2.
\label{eq:quad_dist}
\ee
This is an integral of $P^{\mathrm{eff}}_\DM(\alpha)$ against a slowly varying function of $\alpha$, implying quantum effects are suppressed.
A simple quadrature quantum quantifier is the squeezing parameter $S = \mathrm{var}(X) - 1/2$, which can only be negative for nonclassical states.
Taking moments of Eq.~\eqref{eq:quad_dist} yields 
\be
S_{\mathrm{cav}} = \eta \, S_\DM \geq - \frac{\eta}{2}
\label{eq:S_cav}
\ee
where the minimum is achieved by an infinitely squeezed DM state.
Thus, in the ideal case, DM can drive the cavity only into a very slightly squeezed state.

Similarly, the cavity number distribution is
\be
p_n = \int d \alpha \, P^{\mathrm{eff}}_\DM(\alpha) \, |\la n | \sqrt{\eta} \, \alpha \ra|^2.
\label{eq:num_dist}
\ee
A relevant nonclassicality measure is the Mandel $Q$-parameter, $Q = \mathrm{var}(n)/\la n \ra - 1$, where $Q < 0$ indicates nonclassical sub-Poissonian fluctuations.
Using Eq.~\eqref{eq:num_dist},
\be
Q_{\mathrm{cav}} = \eta \, Q_\DM \geq - \eta
\label{eq:Q_cav}
\ee
where the minimum is achieved by a DM Fock state.
Sub-Poissonian fluctuations in the cavity are thus strongly suppressed.
Intuitively, the low probability of axion-photon conversion imprints approximately Poisson fluctuations on the cavity number.

Observing $Q_{\mathrm{cav}} = - \eta$ is much harder than discovering the axion.
It requires measuring for time $t_{\mathrm{int}} \sim \, t_m / \eta^2$, of order $10^{30}\,$years for the parameters in Eq.~\eqref{eq:eta_value}, and prohibitively long even for the largest possible values of $g_{a\gamma\gamma}$ and $Q_c$.
We derive this scaling in the End Matter, although it follows intuitively as the uncertainty on $Q$ decreases proportional to $\sqrt{t_{\mathrm{int}}/t_m}$, so to obtain an uncertainty below $\eta$ we require the stated scaling.
(We also demonstrate in the End Matter that cavity thermal noise leads to $S_{\mathrm{cav}} \simeq \eta \, S_\DM + n_t$ and $Q_{\mathrm{cav}} \simeq \eta \, Q_\DM + 2 n_t$, consistent with the argument above Eq.~\eqref{eq:T_requirement}.)

Similar reasoning shows that a wide variety of nonclassicality measures are also suppressed.
Nonclassical signatures in the Wigner function~\cite{PhysRevLett.124.133601,PhysRevResearch.3.043116} are suppressed because the Wigner function is a Gaussian convolution of the $P$-function.
Furthermore, nonclassicality measures involving number can often be expressed using the cavity's moment generating function~\cite{PhysRevA.31.338,PhysRevA.41.1721,PhysRevA.41.1569,KLYSHKO19967},
\be
M(\mu) = \sum_{n=0}^{\infty} p_n(1 - \mu)^n
=\int d \alpha\, P^{\mathrm{eff}}_\DM(\alpha)\, e^{-\eta \mu |\alpha|^2}
\ee
which is again an integral of $P^{\mathrm{eff}}_\DM(\alpha)$ against a slowly varying function, washing out effects of negativity.
We expect similar arguments apply to even more general nonclassicality measures involving, e.g.~higher-order quadrature moments~\cite{PhysRevA.46.485,AGARWAL1993109,PhysRevLett.89.283601,PhysRevA.71.011802,PhysRevA.72.043808,PhysRevA.92.011801}.

We emphasize that our result only obstructs measuring the negativity of $P^{\mathrm{eff}}_\DM$.
This implies that for detection, the axion can be treated as a classical mixture of coherent states as in Eq.~\eqref{eq:GSP}, or equivalently as a stochastic classical field, which guarantees $Q_{\mathrm{cav}}, S_{\mathrm{cav}} \geq 0$ (see End Matter).
However, experiments could distinguish different classical mixtures.
For instance, a single-mode coherent DM state would have $S_\DM = Q_\DM = 0$, while the Gaussian state expected from virialization would have $S_\DM \simeq Q_\DM \simeq N_{\mathrm{eff}} \gg 1$.
This would manifest as $S_{\mathrm{cav}}$ and $Q_{\mathrm{cav}}$ scaling as the mean number of signal photons $n_s = \eta N_{\mathrm{eff}}$. 
Thus, if the axion can be detected at all, implying that $n_s$ is not too small, we expect coherent and Gaussian states could be readily distinguished in a post-discovery scenario.
For further details, see Refs.~\cite{Bao:2026prd,Bao:2026kbx}.

\vspace{0.2cm}
\noindent {\bf Entanglement Signatures.}
%
One might object that exotic DM states can produce distinctive entanglement signatures.
For instance, if the DM and cavity start in a cat and vacuum state, $\propto (\ket{\alpha} + \ket{-\alpha})_\DM \ket{0}_{\mathrm{cav}}$, then they evolve to an entangled state, proportional to
\be
\hspace{-0.1cm}\ket{\alpha \cos gt}_\DM \ket{\alpha \sin gt}_{\mathrm{cav}} \!+\! \ket{-\alpha \cos gt}_\DM \ket{-\alpha \sin gt}_{\mathrm{cav}}
\ee
as was noted in Ref.~\cite{Allali:2021puy}.
However, in practice, we can only directly measure the cavity state, so the DM state must be traced out.
In the reduced density matrix of the cavity, off-diagonal terms between $\ket{\alpha \sin gt}$ and $\ket{-\alpha \sin gt}$ are suppressed by the overlap $\la \alpha \cos gt | {-} \alpha \cos gt \ra \simeq \la \alpha | {-} \alpha \ra \sim \exp(- 2 N_{\mathrm{eff}})$.
Thus, since the conversion efficiency is small, the final cavity state is indistinguishable from the more mundane case where DM begins in a classical mixture of $\ket{\alpha}$ and $\ket{-\alpha}$.

It is also interesting to consider decoherence as a signal, as has been suggested for particle DM in Refs.~\cite{Riedel:2012ur,Riedel:2016acj}.
In this case, one would prepare the cavity in a superposition, and observe the loss of purity of the cavity state.
Though it is unclear whether this would be a practical method for detection, this signal is fully described by Eq.~\eqref{eq:final_Pcav}, and our estimates suggest that exotic DM states do not generate decoherence any faster than a Gaussian state.
(Conversely, the quantization of the axion could be established by an entanglement witness~\cite{Rufo:2025rps}.
However, this is very difficult, even compared to the gravitational case, because the axion only mediates dipole forces.)

Finally, our formalism simply generalizes to multiple cavities.
Widely spatially separated cavity modes $c_i$ couple to independent effective modes $a_{\mathrm{eff}, i}$, mixing as in Eq.~\eqref{eq:conj_result}.
Nonclassical effects appear through the negativity of the joint $P$-function $P_{\mathrm{cav}}^f(\alpha_{\mathrm{cav},1}, \ldots, \alpha_{\mathrm{cav},n})$, which accounts for cavity entanglement.
However, the arguments above go through qualitatively unchanged, and show that nonclassical effects remain suppressed by $\eta$.

\vspace{0.2cm}
\noindent {\bf Discussion.}
%
While the quantum mechanics of the detector could be essential in the discovery of DM, the quantum mechanics of DM will not.
Building on the quantum description of DM developed in Ref.~\cite{Cheong:2024ose}, we exhibited two barriers to observing nonclassical effects.
First, if the DM modes are independent, a detector that cannot resolve the fundamental DM modes instead sees a classical effective Gaussian mode by the quantum central limit theorem.
Second, nonclassical effects are always much harder to observe than the DM itself, even in the absence of noise.
We emphasize that neither of these arguments relied on the large occupation number of wave DM.
In short, a classical description of axion DM suffices even if it is in a highly nonclassical state.

Our work resolves a fundamental issue in the axion literature.
The original works on axion DM~\cite{Abbott:1982af,Preskill:1982cy} suggested the axion can be treated classically because it is in a pure coherent state, or because it has large occupation number. 
Neither point is correct, yet similar comments persist in recent authoritative reviews~\cite{Jaeckel:2010ni,Marsh:2015xka,Battaglieri:2017aum,Irastorza:2018dyq}, leading to increasing uncertainty over the validity of the classical field approximation underlying axion experiments.

This confusion has led to claims that axion DM can have inherently quantum signatures.
For example, Ref.~\cite{Marsh:2022gnf} compared coherent and Gaussian DM states, and Ref.~\cite{Lentz:2025lkg} discussed squeezed DM states but considered measuring positive $S_{\mathrm{cav}}$.
However, these signatures can all be realized by a classical axion field.
It has also been claimed that ultralight DM produces strong effects by Bose enhancement~\cite{PhysRevLett.132.091801}, but these effects also appear classically~\cite{Zhou:2025wax}.
Of course it is important to determine the expected state of axion DM in our galaxy, and in particular, if there are strong correlations between modes.
However, given a DM state, detector calculations can be simplified by treating the axion as an appropriate classical ensemble.
We have shown this is a universal consequence of the weakness of wave DM couplings.

Our conclusions generalize broadly to wave DM searches and further to nonclassical signatures of weakly coupled physics.
Indeed, our analysis is intimately related to the classic question of whether one can infer the existence of the graviton from gravitational radiation~\cite{Dyson:2013hbl,Rothman:2006fp,Boughn:2006st,Carney:2023nzz,Carney:2024dsj,Manikandan:2024fmf}.
In that case, it has been established that it is possible to convert gravitons to single quanta and detect those quanta, but genuinely quantum effects that would exclude a classical description of the gravitational field are suppressed by the low conversion efficiency.
Our conclusions go beyond existing results, demonstrating a more general obstruction to the observation of uniquely quantum effects.

Returning to DM, our arguments generalize straightforwardly to any interaction where DM directly converts to a photon.
They also generalize to couplings to matter which produce forces on macroscopic oscillators, with mechanical modes replacing cavity modes.
Two types of signature differ more substantially.
First, the axion can exert torques on spins, whose nonclassical states are described differently (e.g.~see Refs.~\cite{Boyers:2025qgc,delCastillo:2025qnr,Galanis:2025amc}).
Second, dilaton DM shifts the frequency of oscillators, corresponding to $H_{\mathrm{int}} \simeq g \, c^\dagger c \, (a_{\mathrm{eff}} + a_{\mathrm{eff}}^\dagger)$.
Though observing nonclassical effects in both cases should remain difficult, it would be interesting to quantify this explicitly. \\

\medskip
\noindent {\it Acknowledgments.} We thank Daniel Carney, Andrew Eberhardt, Sebastian Ellis, Anson Hook, Giacomo Marocco, David Marsh, Gilad Perez, Ryan Plestid, and Ritoban Basu Thakur for discussions.
YB and LTW are supported by the Department of Energy grant DE-SC0013642.
DYC is supported by the Enrico Fermi and KICP fellowship from the Enrico Fermi Institute and the Kavli Institute for Cosmological Physics at the University of Chicago, and the National Research Foundation of Korea (NRF) grant funded by the Korea government (MSIT) (RS-2024-00340153).
The research of NLR, JT, and KZ was supported by the Office of High Energy Physics of the U.S. Department of Energy under contract DE-AC02-05CH11231.
The work of JT was further supported by the NSF Graduate Research Fellowship Program under Grant DGE2146752.

\bibliographystyle{utphys}
\bibliography{refs}

\clearpage
\appendix
\onecolumngrid
\section*{End Matter}
\twocolumngrid

\setcounter{equation}{0}
\renewcommand{\theequation}{E\arabic{equation}}

\vspace{0.2cm}
\noindent {\bf Properties of the Effective Mode.}
%
Here we justify the scalings of $\Omega$ and $\la a^\dagger_{\mathrm{eff}} a_{\mathrm{eff}} \ra$ quoted below Eq.~\eqref{eq:Hint_def}.
First, normalization of $a_{\mathrm{eff}}$ requires
\bea
\Omega &= \frac{V_c}{4{\cal V}} \sum_{\bp} |C_{\bp}|^2 \\
&= \frac{1}{\cal V} \sum_{\bp}  \frac{m_\DM}{4\omega_p}\,  \bigg|\int_{V_c} \hspace{-0.15cm} d^3 \bx\,  e^{i \bp \cdot \bx} \tilde{E}_z(\bx) \bigg|^2\!.
\eea
To estimate the expression, note that if the cavity has a length $L$, the integral becomes constant for $p L \lesssim 1$ and is rapidly suppressed for $p L \gtrsim 1$.
Thus, the number of modes significantly contributing to the sum is of order $\mathcal{V} / L^3 \sim \mathcal{V} / V_c$.
Since haloscopes have $m_\DM L \sim 1$, the relevant modes have $\omega_p \simeq m_\DM$ and $\bp \simeq \bm{0}$.
Therefore,
\be
\Omega \simeq \frac{1}{4} |C_{\bm{0}}|^2 \sim 1
\label{eq:Omega-estimate}
\ee
where the final scaling holds as $|C_{\bm{0}}|^2$ is the cavity form factor.
(As an explicit example, the TM$_{010}$ mode of a cylindrical cavity has $|C_{\bm{0}}|^2 \simeq 0.69$.)

Next, consider the mean occupancy of $a_{\mathrm{eff}}$,
\bea
\la a_{\mathrm{eff}}^\dagger a_{\mathrm{eff}} \ra &= \frac{V_c}{4 {\cal V}\Omega} \sum_{\bp, \bq} C_{\bp}^* C_{\bq} \, e^{- i (K_q - K_p) t} \la a_{\bp}^\dagger a_{\bq} \ra \\
&\simeq \frac{V_c}{{\cal V}} \sum_{\bp} \frac{|C_{\bp}|^2}{|C_{\bm{0}}|^2} \, \la a_{\bp}^\dagger a_{\bp} \ra.
\eea
The second line used Eq.~\eqref{eq:Omega-estimate} and neglected the momentum cross terms, which vanish on average due to the random phase $e^{- i (K_q - K_p) t}$.
For nonrelativistic DM, the occupied modes have $p \sim m_\DM v_\DM \ll m_\DM$, and in this range $|C_{\bp}|^2$ is roughly constant and equal to $|C_{\bm{0}}|^2$.
The sum then reduces to the number of axions in the quantization region, $(\rho_\DM/m_\DM) {\cal V}$, leaving
\be
\la a_{\mathrm{eff}}^\dagger a_{\mathrm{eff}} \ra \simeq \frac{\rho_\DM V_c}{m_\DM}.
\ee

\vspace{0.2cm}
\noindent {\bf Derivation of Eq.~\eqref{eq:final_Pcav}.}
%
We begin with an initially unentangled DM-cavity state,
\be
\hspace{-0.1cm}\hat{\rho}(0) = \int d \alpha \, d\beta \, P^{\mathrm{eff}}_\DM(\alpha)\, P_{\mathrm{cav}}^i(\beta) \, \ket{\alpha} \bra{\alpha}_\DM\, \ket{\beta} \bra{\beta}_{\mathrm{cav}}.
\ee
The evolution of the cavity density matrix follows by evolving the states as discussed below Eq.~\eqref{eq:conj_result} and tracing out the unobserved final DM state, yielding
\bea
\hat{\rho}_{\mathrm{cav}}(t) = &\int d \alpha \, d\beta \, P^{\mathrm{eff}}_\DM(\alpha) P_{\mathrm{cav}}^i(\beta) \\ 
\times\, &\ket{\beta \cos gt + \alpha \sin gt} \bra{\beta \cos gt + \alpha \sin gt}.
\eea
To identify $P_{\mathrm{cav}}^f$, we change variables to $\alpha' = \alpha \sin gt + \beta \cos gt$ and $\beta' = \alpha \sin gt$ and drop the primes, to find
\bea
\hat{\rho}_{\mathrm{cav}}(t) = &\int \frac{d \alpha \, d\beta}{\sin^2(gt) \, \cos^2(gt)}  \\
\times\, & P^{\mathrm{eff}}_\DM \bigg( \frac{\beta}{\sin gt} \bigg) P_{\mathrm{cav}}^i\bigg( \frac{\alpha - \beta}{\cos gt} \bigg) \, \ket{\alpha} \bra{\alpha}.
\eea
Setting $t = t_m$ and introducing $\eta$, we arrive at Eq.~\eqref{eq:final_Pcav}.
Similar reasoning can be used to derive Eq.~\eqref{eq:Peff_def}.

\vspace{0.2cm}
\noindent {\bf Illustrating Nonclassical $P$-functions.}
%
Visualizing nonclassical $P$-functions can be difficult, as for Fock and cat states the $P$-function is a highly singular distribution~\cite{LMandel_1986}.
However, this can be an artifact of considering overly idealized states.
For example, a Fock state with an arbitrarily small amount of Gaussian noise has a rapidly falling normal-ordered characteristic function, whose Fourier transform is a well-behaved $P$-function.
Thus, for illustration in Fig.~\ref{fig:3d_plot}, we take the DM state to be that of adding $7$ quanta to a thermal background of $1.2$ quanta, using the $P$-function derived in Ref.~\cite{PhysRevA.46.485}.

\vspace{0.2cm}
\noindent {\bf Extension to Lower Mass Axions.}
%
The fact that the efficiency in Eq.~\eqref{eq:eta_value} scales as $\eta \propto m_\DM^{-2}$ superficially suggests that nonclassical effects could be easier to observe for lower masses.
This is not correct.
Firstly, at lower masses, the requirement in Eq.~\eqref{eq:T_requirement} on the temperature becomes much more stringent.
Furthermore, when $m_\DM L \ll 1$, a haloscope with a static background field has a form factor falling as $C_{\bm{0}} \sim m_\DM L$~\cite{DMRadio:2022pkf}, so that the conversion efficiency actually scales as
\be
\eta \sim g_{a\gamma\gamma}^2 B_0^2 (m_\DM L)^2 (Q_c/m_\DM)^2
\ee
which is not enhanced for lower $m_\DM$.
By contrast, in the ``heterodyne'' approach with an oscillating background field, the form factor remains order-one~\cite{Li:2025pyi}, but the temperature is much higher, $T \simeq 2 \, \mathrm{K}$.
Thus, in either case detecting quantum effects remains very difficult.

\vspace{0.2cm}
\noindent {\bf Time to Observe Mandel $Q$.}
%
For simplicity, consider a haloscope at the edge of detection: in most measurements there are no signal photons, so $n_s = \eta N_{\mathrm{eff}} \ll 1$.
Then $p_0 \simeq 1$ and discovering the axion corresponds to measuring a nonzero value of $p_1 \simeq n_s$.
Mandel $Q$ imprints itself on the even smaller probability $p_2$, as
\be
p_2 \simeq \frac{1}{2} \, n_s (n_s + Q_{\mathrm{cav}}).
\ee
We can therefore infer $Q$ through a measurement of $p_2$; the difference in $p_2$ between the minimum allowed classical ($Q_{\mathrm{cav}} = 0$) and quantum ($Q_{\mathrm{cav}} = -\eta$) values is $\Delta p_2 \simeq n_s \eta/2$.
If one performs $N_{\mathrm{shot}} \gg 1$ measurements, the expected number of events with $2$ photons is $p_2 N_{\mathrm{shot}}$, which provides an estimator for $p_2$.
The variance of our estimator is roughly $p_2/ N_{\mathrm{shot}} \simeq n_s^2/2N_{\mathrm{shot}}$ and requiring this to be less than $(\Delta p_2)^2$, we conclude $N_{\mathrm{shot}} \sim 1/\eta^2$ measurements are required.
Therefore, measuring an inherently quantum $Q$ requires a total integration time $t_{\mathrm{int}} \sim t_m / \eta^2$, as stated in the main text.
(This scaling persists if we estimate $Q$ directly, rather than $p_2$, and remove the $n_s \ll 1$ assumption.)

Quantitatively, this time is exceptionally long,
\bea
\frac{t_m}{\eta^2} \sim 10^{30} \, \mathrm{years} \left( \frac{10^{-15} \, \mathrm{GeV}^{-1}}{\ga} \frac{10 \, \mathrm{T}^{\vrule height 4.5pt width 0pt}}{B_0} \right)^4 \\
\times \left( \frac{10^5}{Q_c} \frac{m_\DM}{10^{-5} \, \mathrm{eV}} \right)^3\!.
\eea
Even if we consider maximally optimistic values for a superconducting cavity, $g_{a\gamma\gamma} = 10^{-12} \, \mathrm{GeV}^{-1}$,  $B_0 = 0.2 \, \mathrm{T}$ and $Q_c = 10^{12}$, the integration time only reduces to $10^4 \, \mathrm{years}$, still prohibitively long.

It is far easier to distinguish positive values of Mandel $Q$, as they can span a much larger range.
For a coherent DM state $Q_{\mathrm{cav}} = 0$, whereas a Gaussian DM state yields $Q_{\mathrm{cav}} = n_s$.
Again, taking $n_s \ll 1$, these correspond to $p_2 \simeq n_s^2/2$ and $p_2 \simeq n_s^2$ respectively, consistent with the number distribution being Poisson or geometric.
(A similar point was discussed in Ref.~\cite{Manikandan:2024fmf}.)
Distinguishing these scenarios requires only an order-one fractional uncertainty on $p_2$, $\Delta p_2 \sim n_s^2$, and therefore $N_{\mathrm{shot}} \sim 1/n_s^2$ and $t_{\mathrm{int}} \sim t_m / n_s^2$.
This is longer than the time $t_{\mathrm{disc}} \sim t_m / n_s$ that would be required to discover the axion, but if the axion can be discovered in a scanning search, then the difference between them is at most several orders of magnitude.
Thus, distinguishing positive values of $Q_{\mathrm{cav}}$ is feasible in a post-discovery scenario.

\vspace{0.2cm}
\noindent {\bf $S_{\mathrm{cav}}$ and $Q_{\mathrm{cav}}$ with Thermal Noise.}
%
Here we recompute Eqs.~\eqref{eq:S_cav} and \eqref{eq:Q_cav} in the presence of cavity thermal noise.
We work in the Heisenberg picture and a rotating frame to remove factors of $e^{-im_\DM t}$.
The cavity mode then evolves as
\be
c(t_m) = \sqrt{1-\eta}\, c_0 + \sqrt{\eta}\, a_0
\label{eq:ctm}
\ee
where $c_0$ and $a_0$ are the initial operators.
(This should be contrasted with Eq.~\eqref{eq:conj_result}, which was not the time evolution of $c$, but rather an identity used to infer the time evolution of the states in the interaction picture.)
Now, the desired quantities can be expressed as 
\bea
S_{\mathrm{cav}} &= \frac{1}{2} \Big( \la \normord{(c + c^\dagger)^2} \ra - \la c + c^\dagger \ra^2 \Big) \\
Q_{\mathrm{cav}} &= \frac{\la c^\dagger c^\dagger c c \ra-\la c^\dagger c \ra^2}{\la c^\dagger c \ra}
\label{eq:SQ}
\eea
with a pair of colons denoting normal ordering.

To evaluate these expectation values, we note that for an initial thermal cavity state, $\la c_0^\dagger c_0 \ra = n_t$ and $\la c_0^\dagger c_0^\dagger c_0 c_0 \ra = 2 n_t^2$, with the remaining relevant expectation values vanishing.
We then find
\be
S_\mathrm{cav} = \eta S_\DM + \bar{n}_t 
\ee
with $\bar{n}_t = (1 - \eta) n_t$.
Similarly,
\bea
Q_\mathrm{cav} &= \frac{\eta n_s Q_\DM + \bar{n}_t (2 n_s + \bar{n}_t)}{n_s + \bar{n}_t} \\
&\simeq \eta Q_\DM + 2 \bar{n}_t
\eea
where the final expression holds for $n_t \ll n_s$, the relevant limit for measuring small $Q_{\mathrm{cav}}$.
Given $\bar{n}_t \simeq n_t$ this justifies the expressions in the main text.

In principle, even if we measured a positive value for $Q_{\mathrm{cav}}$ or $S_{\mathrm{cav}}$, we could use these results to infer a negative (nonclassical) result for DM, but only if $n_t$ was known to an absolute precision of ${\sim} \eta$.
This would be very difficult.
Furthermore, non-thermal noise sources such as detector dark counts would also effectively contribute to $n_t$, and must be known to the same precision.
Finally, we cannot avoid this problem by preparing the cavity in a nonclassical initial state, as this would still require controlling the cavity to ${\sim} \eta$ precision.

\vspace{0.2cm}
\noindent {\bf Classical DM Yields $Q,S \geq 0$.}
%
If DM is a classical field, it cannot drive an initially empty cavity to have a negative value for $S_{\mathrm{cav}}$ or $Q_{\mathrm{cav}}$.
We can see this as follows.
If DM is a classical field, then Eq.~\eqref{eq:ctm} becomes $c(t_m) \sim c_0 + \sqrt{\eta}\, \alpha$, where $\alpha$ is a c-number, which in general can be a classically random variable.
Substituting this into Eq.~\eqref{eq:SQ} and assuming the cavity is initially in vacuum, we have
\be
S_{\mathrm{cav}}^{\textrm{cl.}} = 2\eta \, \textrm{var} \big( \textrm{Re}\, \alpha \big),\hspace{0.2cm}
Q_{\mathrm{cav}}^{\textrm{cl.}} = \eta \, \textrm{var}\big(|\alpha|^2\big)/\la |\alpha|^2 \ra
\ee
where in both cases the variance and expectation value are taken over the classical probability distribution for $\alpha$.
As the variance is positive semidefinite due to the Cauchy--Schwarz inequality, we conclude $S^{\textrm{cl.}}_{\mathrm{cav}}, Q^{\textrm{cl.}}_{\mathrm{cav}} \geq 0$.

The above is an example of a more general phenomenon originally noted by Glauber~\cite{Glauber:1963tx}: classical random currents drive a coupled quantum system into a classically random ensemble of coherent states.
Indeed, $c \sim c_0 + \sqrt{\eta} \,\alpha = D^{\dag}(\sqrt{\eta}\alpha) \,c_0\, D(\sqrt{\eta}\alpha)$, with $D(\alpha) = \exp \!\big(\alpha c_0^{\dag} - \alpha^* c_0\big)$ the displacement operator.
Whilst suggestive, we can formalize this further.
For a classical axion field, the Hamiltonian in Eq.~\eqref{eq:Hint_def} could be written $H_{\mathrm{int}} = i g \big(c^{\dag}  \alpha_{\mathrm{eff}} - c \alpha_{\mathrm{eff}}^* \big)$.
The interaction-picture time-evolution operator is then,
\begin{align}
U_I(t) = \,&\exp \!\Bigg( {-} i \int_0^t \!d\tau\,H_{\mathrm{int}} \Bigg) 
= \exp \!\big(gt(c^{\dag} \alpha_{\mathrm{eff}} - c \alpha_{\mathrm{eff}}^*) \big) \nonumber \\
\simeq \,&D\big(\sqrt{\eta}\, \alpha_{\mathrm{eff}}\big)
\end{align}
where we took $gt \simeq \sqrt{\eta}$ in analogy with Eq.~\eqref{eq:eta_value}.
Therefore, for a fixed $\alpha_{\mathrm{eff}}$ time evolution displaces the cavity $P$-function to $P_{\mathrm{cav}}(\alpha-\sqrt{\eta}\, \alpha_{\mathrm{eff}})$; the classical driver simply shifts the location of the distribution.
The extension to include time dependence or classical randomness in $\alpha_{\mathrm{eff}}$ is straightforward, and in all cases the classical source does not generate negativity in $P_{\mathrm{cav}}(\alpha)$.

\clearpage

\end{document}